\documentclass[aip,amsmath,amssymb,reprint]{revtex4-1}

\usepackage{graphicx}
\usepackage[utf8]{inputenc}
\usepackage[T1]{fontenc}
\usepackage{mathptmx}
\usepackage{xcolor}
\usepackage[normalem]{ulem}  
\usepackage[colorlinks=true,linkcolor=blue,citecolor=blue,urlcolor=blue]{hyperref}%

\definecolor{orange}{rgb}{1,0.5,0}
\definecolor{darkgreen}{rgb}{0.13, 0.55, 0.13}

\makeatletter
\newcommand{\mathleft}{\@fleqntrue\@mathmargin0pt}
\newcommand{\mathcenter}{\@fleqnfalse}
\makeatother

\newcommand\commentout[1]{}

\begin{document}

\title{Thermoelectric study of the time-dependent Resonant Level Model}

\author{Adel Kara Slimane}
\author{Genevi\`eve Fleury}
\affiliation{Universit\'e Paris-Saclay, CEA, CNRS, SPEC, 91191 Gif-sur-Yvette, France}

\begin{abstract}
We study the non-interacting time-dependent resonant level model mimicking a driven quantum dot connected through leads to two electronic reservoirs held at different temperatures and electrochemical potentials. Using a scattering approach, we provide analytical formulas of the time-dependent particle currents, heat currents, and input driving power under the wide-band limit approximation. We also derive Landauer formulas for the corresponding time-integrated quantities when the perturbation applied on the dot is of finite duration. Then, we focus on the case of a single square pulse, benchmark our analytical results against numerical ones that are valid beyond the wide-band limit, and perform numerical simulations for a smooth square pulse and a periodic square pulse train.
Finally, we discuss whether the efficiency of the device in a stationary Seebeck configuration can be enhanced by driving the dot potential. We find numerically that the transient increase of the efficiency observed in some cases is eventually cancelled out at long times.
\end{abstract}

\maketitle

\section{Introduction}

Molecular junctions and quantum dots have emerged as promising experimental platforms for investigating energy transport on the nanoscale. They have been used in the last few years to investigate thermoelectric effects\cite{Dzurak1997,Scheibner2005,Rincon2016,Cui2017,Cui2018,Prete2019,Dutta2019} and to implement \textit{e.g.} heat engines\cite{Josefsson2018,Josefsson2020,Jaliel2019,Popp2020}, heat valves\cite{Dutta2020}, and thermal rectifiers\cite{Scheibner2008} in the stationary regime. The physics in those devices is often described by invoking the paradigmatic fermionic Resonant Level Model (RLM): a single localized electronic level with energy $\epsilon_0$ (hereafter called the dot) is connected with coupling energies $\Gamma_\alpha$ to two (or more) electronic baths $\alpha$ held at temperatures $T_\alpha$ and electrochemical potentials $\mu_\alpha$. 
While it is straightforward to calculate heat and charge currents for the non-interacting RLM in the stationary regime, the inclusion of electron-electron interactions in the dot and/or of time-dependent perturbations pushing the system out-of-equilibrium complicates considerably the model.\\
\indent In particular, in spite of its apparent simplicity, the non-interacting time-dependent RLM is the subject of active research in the fields of high-frequency nanoelectronics and quantum thermodynamics. It has been extensively used in the theoretical literature to investigate dynamical charge\cite{jauho1994,Platero2004,Gurvitz2015,ridley2015,Gurvitz2021}  and energy\cite{esposito2010,crepieux2011, Liu2012, esposito2015, ludovico2016bis, zhou2015, dare2016, yu2016, Entin2017, Lehmann2018, covito2018, haughian2018,bruch2018,Oz2020,Semenov2020,Yu2020} transport with different techniques. Fundamental questions have been addressed with the aim of building a consistent theory of thermodynamics in the out-of-equilibrium quantum regime\cite{ludovico2016bis,haughian2018,bruch2018,Oz2020,Semenov2020}. The possibility of performing thermodynamic tasks such as heat pumping\cite{cuansing2020}, cooling\cite{dare2016} or heat-to-work conversion\cite{crepieux2011, zhou2015, Yu2020} has also been investigated. In particular, in Ref.\cite{zhou2015}, the authors reported on a boost of thermoelectric efficiency of a driven heat engine modeled by the time-dependent RLM. More advanced models\cite{Rey2007,Juergens2013,Bhandari2020,Harunari2021,Mayrhofer2021} have been considered as well, \textit{e.g.} by including Coulomb interaction and/or by describing a two-level dot or two dots instead of one.\\ 
\indent In this paper, we study thermoelectric transport in the non-interacting RLM when the energy $\epsilon_0$ of the dot is made time-dependent. As in Refs.\cite{bruch2018,covito2018,kara2020}, the dot is coupled from the remote past to two baths through ideal (non-interacting) leads. Dissipation only occurs in the baths and a static temperature and/or electrochemical potential bias can be applied between the two baths. We use a scattering approach\cite{Michelini2019,kara2020} to describe time-dependent thermoelectric transport in our model. This approach is not restricted to weak dot-bath couplings and is valid for arbitrary driving beyond the adiabatic limit. It is only assumed that the driving is switched on from a given instant $t=0$ (\textit{i.e.} $\epsilon_0(t<0)=V_0$ constant) so that the time-dependent scattering state inside the dot and the leads can be calculated\cite{gaury2014a} by evolving in time the stationary scattering state defined for $t<0$. In Ref.\cite{kara2020}, we used this approach to develop a numerical technique\cite{tKwantop} -- based on the Tkwant software\cite{kloss2021,tKwant} -- for simulating time-dependent thermoelectricity in quantum systems. We used the driven RLM as a benchmark and performed numerical simulations of dynamical Peltier cooling in a two-dimensional quantum point contact. Our results pointed to a negative role of time-dependent perturbations for thermoelectric cooling but did not allow us to comprehend and disentangle the physical effects at stake. 
Here, we focus our interest on the time-dependent RLM and leverage our scattering approach to provide an analytical description of time-dependent thermoelectric transport within the so-called wide-band limit (WBL) approximation. Our analytics serve as guidelines for identifying interesting operating regimes that can be investigated subsequently numerically. 
In particular, inspired by Ref.\cite{zhou2015}, we ask ourselves whether or not the efficiency of the stationary device in a Seebeck configuration can be enhanced by driving the dot. We discuss the subtleties of defining a time-resolved efficiency and draw the empirical conclusion from the analysis of our numerous numerical data that the time-dependent driving does not bring an advantage in terms of net thermodynamic efficiency in the RLM. Even though the efficiency can be increased in the transient regime, the net efficiency at long times 
does not exceed the stationary efficiency in the investigated wide parameter range.\\
\indent The outline of the paper is as follows. We introduce in Sec.\ref{sec_model} our time-dependent RLM and in Sec.\ref{sec_scattth} the basics of the time-dependent scattering theory of thermoelectric transport. This theory is used in Sec.\ref{sec_analytics_WBL} within the WBL approximation to derive (semi-)analytical formulas describing time-dependent thermoelectric transport in the RLM. Sec.\ref{sec_sq} is devoted to the peculiar case where a single square pulse is applied on the dot. Potential applications for time-dependent thermoelectric energy harvesting are discussed in Sec.\ref{sec_eff}. We conclude in Sec.\ref{sec_ccl}.

\section{Model}
\label{sec_model}
\begin{figure}[t]
    \centering
    \includegraphics[keepaspectratio,width=0.9\columnwidth]{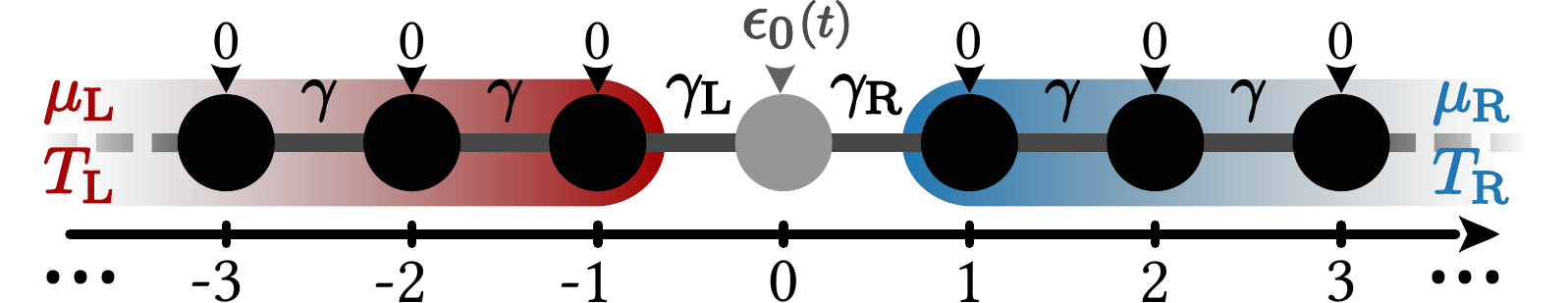}
    \caption{Sketch of the time-dependent Resonant Level Model studied in this paper. A central site playing the role of the dot is attached to two left ($L$) and right ($R$) leads modeled by semi-infinite 1D chains. For times $t\leq 0$, the model Hamiltonian is static but the system can be driven out of equilibrium by applying a temperature bias $T_L-T_R$ and/or an electrochemical potential bias $\mu_L-\mu_R$ across the left and right electronic reservoirs attached to the leads 
    Additionally, for times $t>0$, the onsite energy $\epsilon_0$ in the dot can be varied in time with \textit{e.g.} a back gate. }
    \label{fig_RLM_sys}
\end{figure}

We consider a one-dimensional (1D) discretized version of the time-dependent RLM sketched in Fig.\ref{fig_RLM_sys}. A central site $0$ with time-dependent onsite energy $\epsilon_0(t)$ is attached through nearest-neighbor hopping terms $\gamma_L$ and $\gamma_R$ to two left ($L$) and right ($R$) semi-infinite 1D chains playing the role of the leads. In the latter, the nearest-neighbor hopping term is denoted by $\gamma$ and the onsite energies are taken equal to zero for simplicity. The tight-binding Hamiltonian of our system reads
\begin{equation}
    \label{eq_RLM_1D_1}
    H(t)=H_0(t)+\sum_{\alpha=L,R}H_\alpha+\sum_{\alpha=L,R}H_{0\alpha}
\end{equation}
where $H_0(t)=\epsilon_0(t)c_0^\dagger c_0$ is the dot Hamiltonian, $ H_L=-\gamma\sum_{i\leq  -1}c_{i- 1}^\dagger c_{i}+h.c.$ and $ H_R=-\gamma\sum_{i\geq 1}c_{i+1}^\dagger c_{i}+h.c.$ are the Hamiltonians of the left and right leads respectively, while $H_{0L} =-\gamma_L \,c_0^\dagger c_{-1}+h.c.$ and $H_{0R} =-\gamma_R \,c_0^\dagger c_{1}+h.c.$ are the tunneling Hamiltonians between the dot and the leads. Here, $c_i$ and $c_i^\dagger$ denote the annihilation and creation operators of an electron at site $i$. Moreover, each lead $\alpha$ ($=L$ or $R$) is attached from the remote past to an electronic reservoir characterized by its (static) electrochemical potential $\mu_\alpha$ and temperature $T_\alpha$. Importantly, we assume that the dot onsite energy is constant for $t\leq 0$ and equal to $\epsilon_0(t\leq 0)=V_0$ while time-dependent perturbations are switched on for $t>0$. We use the notation $\epsilon_0(t)=V_0+V(t)$ with $V(t\leq 0)=0$.

Throughout the paper, we take $e=\hbar=k_B=1$ where $e$ is the electron charge, $\hbar$ the reduced Planck constant, and $k_B$ the Boltzmann constant.

\section{Scattering theory for time-dependent thermoelectric transport}
\label{sec_scattth}
We use a wave-function approach\cite{gaury2014a,Michelini2019,kara2020} to study time-dependent quantum transport in our model. The central objects of the theory are the scattering states $\Psi^{\alpha E}(t)$. 
Since the leads are time-independent, the incoming part of $\Psi^{\alpha E}(t)$ is made of a single plane wave at energy $E$ coming from lead $\alpha$ ($\alpha$ and $E$ being thus used to label the scattering states). 
On the contrary the outgoing part of $\Psi^{\alpha E}(t)$ writes in general as a superposition of outgoing plane waves at different energies $E'$, the dot onsite energy $\epsilon_0(t)$ being time-dependent. For instance, the scattering state $\Psi^{L E}_n(t)$ originating from the left lead and evaluated at site $n$ reads in the leads
\begin{subequations}
\label{eq_PsiLE}
\begin{align}
    \Psi^{LE}_{n<0}(t) &= \chi^{\rightarrow}_n(t,E)+\int\!\frac{\mathrm{d}E'}{2\pi}\chi^{\leftarrow}_n(t,E')\, r(E',E) \\
    \Psi^{LE}_{n>0}(t) &= \int\!\frac{\mathrm{d}E'}{2\pi}\chi^{\rightarrow}_n(t,E')\, d(E',E)
\end{align}
\end{subequations}
where $r(E',E)$ is the probability amplitude for an electron with an energy $E$ coming from the left lead to be reflected with an energy $E'$, $d(E',E)$ is its probability amplitude to be transmitted to the right lead with an energy $E'$, while $\chi^{\rightarrow}_n$ and $\chi^{\leftarrow}_n$ are plane waves propagating in the leads from left to right and right to left respectively \textit{i.e.}
\begin{subequations}
\begin{align}
   \chi^{\rightarrow}_n(t,E)& =\tfrac{1}{\sqrt{|v(E)|}}e^{-iEt+ik(E)n} \\
   \chi^{\leftarrow}_n(t,E)&=\tfrac{1}{\sqrt{|v(E)|}}e^{-iEt-ik(E)n}
\end{align}
\end{subequations}
$v(E)$ and $k(E)$ being the plane wave velocity and momentum satisfying $|v(E)|=\sqrt{4\gamma^2-E^2}$, $E=-2\gamma\cos k(E)$, $k(E)>0$. Similar formula can be written for the scattering state $\Psi^{RE}_{n\neq 0}(t)$ originating from the right lead, with reflection and transmission amplitudes $r'(E',E)$ and $d'(E',E)$. In particular, in the static case,
transport is elastic and $d(E',E)=2\pi\delta(E-E')d_0(E)$, $d'(E',E)=2\pi\delta(E-E')d'_0(E)$, $r(E',E)=2\pi\delta(E-E')r_0(E)$, $r'(E',E)=2\pi\delta(E-E')r'_0(E)$
where $\delta$ is the Dirac distribution and $d_0(E)$, $d_0'(E)$, $r_0(E)$, and $r_0'(E)$ are notations for the scattering amplitudes of the time-independent problem.

Within this framework, the particle current $I^N_L(t)=I^N_{0,-1}(t)$ flowing from the left lead to the dot -- evaluated between the sites $-1$ and $0$ -- and the particle current $I^N_R(t)=I^N_{0,1}(t)$ flowing from the right lead to the dot -- evaluated between the sites $1$ and $0$ -- are given by
\begin{equation}
\label{eq_IN(t)_general}
    I^N_\alpha(t) = 2\gamma_\alpha\!\sum_{\beta}\int\!\frac{\mathrm{d}E}{2\pi}f_\beta(E)\mathrm{Im}[(\Psi^{\beta E}_{\pm 1}(t))^*\Psi^{\beta E}_0(t)] 
\end{equation}
with a sign $+$ [$-$] in the subscript $\pm$ for $\alpha=R$ $[L]$. Here $f_\beta(E)=1/\{1+\exp[(E-\mu_\beta)/T_\beta]\}$ is the Fermi-Dirac distribution of the lead $\beta$ and the sum over $\beta$ is implicitly done over both leads $L$ and $R$. Note that particle number conservation implies
\begin{equation}
    \label{eq_chargecsv}
    I^N_L(t)+I^N_R(t)=\frac{\mathrm{d}\rho_0}{\mathrm{d}t}
\end{equation}
where $\rho_0(t)=\sum_{\beta}\int\!\frac{\mathrm{d}E}{2\pi}f_\beta(E)|\Psi^{\beta E}_0|^2$ is the particle density in the dot.

The corresponding heat currents $I^H_\alpha(t)$ defined by\footnote{Note that the right-hand side of Eq.\eqref{eq_df_IH(t)} is not invariant under a gauge transformation of the time-dependent electromagnetic field. A gauge invariant definition of $I^H_\alpha(t)$ was put forward in Ref.\cite{kara2020}. It reduces to Eq.\eqref{eq_df_IH(t)} for our model \eqref{eq_RLM_1D_1} written in a gauge in which the leads are time-independent.} 
\begin{equation}
    \label{eq_df_IH(t)}
    I^H_\alpha(t)=-\frac{\mathrm{d}}{\mathrm{d}t}\langle H_\alpha+\frac{1}{2} H_{0\alpha} \rangle-\mu_\alpha I^N_\alpha(t)
\end{equation} 
read in terms of the scattering states\cite{kara2020}
\begin{align}
\label{eq_IH(t)_general}
    I^H_\alpha(t) = & \left(\frac{\epsilon_0(t)}{2}-\mu_\alpha\right)I^N_\alpha(t)  + \sum_{\beta}\int\!\frac{\mathrm{d}E}{2\pi}f_\beta(E) \big\{  \nonumber\\ 
    &\mathrm{Im}[(\Psi^{\beta E}_{\mp 1})^*\gamma_L\gamma_R\Psi^{\beta E}_{\pm 1} -(\Psi^{\beta E}_{\pm 2})^*\gamma\gamma_\alpha\Psi^{\beta E}_{0}]\big\} 
\end{align}
where the upper [lower] sign has to be chosen in $\pm$ and $\mp$ for $\alpha=R$ $[L]$.
Finally the input power $P(t)$ corresponding to the driving of the dot energy level  writes 
\begin{equation}
\label{eq_P_df}
    P(t)=-V(t)\frac{\mathrm{d}\rho_0}{\mathrm{d}t}\,.
\end{equation}
Hereafter, we use one exact numerical technique and one approximate semi-analytical technique to calculate $I^N_L(t)$, $I^N_R(t)$, $I^H_L(t)$, $I^H_R(t)$, and $P(t)$.
For numerical simulations, we use the Tkwant software\cite{kloss2021,tKwant} together with its extension package\cite{kara2020,tKwantop} for thermoelectric transport. In brief, Tkwant computes the scattering states $\Psi^{\beta E}(t)$ by solving the time-dependent Schr\"odinger equation for the open system with a Runge-Kutta solver. This requires a subtle treatment of the leads\cite{weston2016a}. Then, the integral over the energy $E$ in Eqs.\eqref{eq_IN(t)_general} and \eqref{eq_IH(t)_general} is done in momentum space with a (Gauss-Kronrod) adaptative scheme. This allows us to compute $I^N_\alpha(t)$, $I^H_\alpha(t)$, and $P(t)$ exactly for a given (arbitrarily high) level of precision and an arbitrary form of $V(t)$. While this numerical approach can help us to explore dynamical thermoelectric transport in the RLM, it is insufficient to provide a physical understanding of the mechanisms at stake. Therefore, to gain more physical insight, we also construct a semi-analytical approach outlined in the next section. 

\section{Semi-analytical treatment in the wide-band limit}
\label{sec_analytics_WBL}
\subsection{Time-resolved formulas}
It is possible to simplify the formulas above under the wide-band limit (WBL) approximation. The WBL is reached in our model \eqref{eq_RLM_1D_1} by increasing\cite{covito2018} the bandwidth $4\gamma$ of the 1D leads, keeping fixed the coupling energies $\Gamma_L\equiv 2\gamma_L^2/\gamma$ and $\Gamma_R\equiv 2\gamma_R^2/\gamma$. In practice, this is achieved by doing the substitutions $\gamma\to\lambda\gamma$, $\gamma_L\to\sqrt{\lambda}\gamma_L$, and $\gamma_R\to\sqrt{\lambda}\gamma_R$, $\lambda$ being a scaling parameter. When $\lambda\to\infty$, the self-energy $\Sigma_\alpha(E)$ of each lead $\alpha$ converges to a pure imaginary energy-independent value, $\Sigma_\alpha(E)\to -i\Gamma_\alpha/2$, and the WBL is reached.

Introducing the Fourier transforms $r(t,E)$ and $d(t,E)$ of the reflection and transmission amplitudes $r(E',E)$ and $d(E',E)$, \textit{e.g.}
\begin{equation}
    r(t,E)=\int\!\frac{\mathrm{d}E'}{2\pi}e^{-iE't}r(E',E)\,,
\end{equation}
 we find in the WBL (see Appendix \ref{appendixA})
\begin{equation}
    \label{eq_IN(t)_1}
    I^N_\alpha(t)=\int\!\frac{\mathrm{d}E}{2\pi}\, \mathcal{I}^N_\alpha(t,E)
\end{equation}
where
\mathleft
\begin{align}
    \mathcal{I}^N_L(t,E)&=f_L(E)[1-|r(t,E)|^2]-f_R(E)|d(t,E)|^2 \label{eq_IN(t)_2}\\
    \mathcal{I}^N_R(t,E)&=f_R(E)[1-|r'(t,E)|^2]\!-\!f_L(E)|d(t,E)|^2 \label{eq_IN(t)_3}
\end{align}
\mathcenter
and 
\begin{align}
     I^H_\alpha(t)=&\int\!\frac{\mathrm{d}E}{2\pi} \Big\{ (E-\mu_\alpha)\, \mathcal{I}^N_\alpha(t,E) \nonumber\\
    & + \left[\frac{\Gamma_\alpha}{\Gamma_{\bar{\alpha}}}f_\alpha(E)+f_{\bar{\alpha}}(E)\right]\left[E|d(t,E)|^2+ \mathrm{Im}\left[d^*\partial_t d\right]\right] \nonumber\\
    & - \sqrt{\frac{\Gamma_\alpha}{\Gamma_{\bar{\alpha}}}}f_\alpha(E)\,\mathrm{Re}\left[\partial_t A(t,E)\right]  \Big\} \label{eq_IH(t)_WBL}
\end{align}
where $A(t,E)= -ie^{iEt}d(t,E)$ and $\bar{\alpha}=R\,[L]$ if $\alpha=L\,[R]$. Note that the alternative\cite{crepieux2011,zhou2015} heat current $\tilde{I}^H_\alpha(t)$ defined by $\tilde{I}^H_\alpha(t)=-\frac{\mathrm{d}}{\mathrm{d}t}\langle H_\alpha\rangle-\mu_\alpha I^N_\alpha(t)$ instead of Eq.\eqref{eq_df_IH(t)} obeys Eq.\eqref{eq_IH(t)_WBL} without the last term in its right hand side. Obviously, $\int\! \mathrm{d}t\, I^H_\alpha(t)=\int\! \mathrm{d}t\, \tilde{I}^H_\alpha(t)$ when $V(t)$ is a pulse of finite duration. Finally, the input power reads
\begin{equation}
    \label{eq_P_general_WBL}
    P(t)=-V(t)\!\int\!\frac{\mathrm{d}E}{2\pi}\!\left[\frac{f_L(E)}{\Gamma_R}+\frac{f_R(E)}{\Gamma_L}\right]\partial_t|d(t,E)|^2\,.
\end{equation}
We point out that our equations for $P(t)$ and $I^N_\alpha(t)$ on one hand, and $I^H_\alpha(t)$ and $\tilde{I}^H_\alpha(t)$ on the other hand, are an equivalent reformulation in the time-dependent scattering approach of the formulas derived in Ref.\cite{jauho1994} and Refs.\cite{crepieux2011,zhou2015,dare2016,kara2020} respectively, within the non-equilibrium Green's function formalism. The bond between the two approaches is provided by the generalized Fisher-Lee formula for time-dependent transport,\cite{gaury2014a} linking the scattering amplitudes to the retarded Green's function. 

\subsection{Time-integrated formulas}
\label{sec_timeintegrated_WBL}
The formula given above are valid under the WBL approximation, with no assumption on the shape of the pulse $V(t)$ in the dot. We now assume that the duration of the pulse $V(t)$ is finite (\textit{i.e.} $V(t)\to 0$ when $t\to\infty$) and introduce the following time-integrated quantities induced by the time-dependent drive
\begin{align}
    \Delta N_\alpha &=\int_0^\infty\!\! \mathrm{d}t\,[I^N_\alpha(t)-I^N_\alpha(t=0)] \\
    \Delta Q_\alpha &=\int_0^\infty\!\! \mathrm{d}t\,[I^H_\alpha(t)-I^H_\alpha(t=0)] \\
    W_{ext} &=\int_0^\infty\!\! \mathrm{d}t\,P(t)
\end{align}
where $\alpha=L$ or $R$. $\Delta N_\alpha$ and $\Delta Q_\alpha$ correspond respectively to the dynamically injected number of particles and heat flowing from lead $\alpha$, while $W_{ext}$ is the driving work. We also define
\begin{equation}
    T_{dyn}(E)=\int_0^\infty\!\!\mathrm{d}t \left[|d(t,E)|^2-|d_0(E)|^2\right]\,.
\end{equation}
In Appendix \ref{appendixA}, we show
\begin{align}
    \label{eq_Tdyn}
    T_{dyn}(E)&= \int_0^\infty\!\!\mathrm{d}t \left[|r_0(E)|^2-|r(t,E)|^2\right]\\
    &= \int_0^\infty\!\!\mathrm{d}t \left[|r_0(E)|^2-|r'(t,E)|^2\right] \label{eq_Tdyn2}
\end{align}
with $\int\!\mathrm{d}E\,T_{dyn}(E)=0$, and
\begin{equation}
    \label{eq_DeltaN}
    \Delta N_L=-\Delta N_R=\int\!\frac{\mathrm{d}E}{2\pi}T_{dyn}(E)[f_L(E)-f_R(E)]
\end{equation}
which can be seen as a generalized Landauer formula as previously noticed in Refs.\cite{gaury2014a,Gurvitz2019}. Moreover, we have
\begin{align}
    \Delta Q_\alpha = & \int\!\frac{\mathrm{d}E}{2\pi}(E-\mu_\alpha)T_{dyn}(E)[f_\alpha(E)-f_{\bar{\alpha}}(E)] \nonumber\\
     & -\frac{\Gamma_\alpha}{\Gamma_\alpha+\Gamma_{\bar{\alpha}}}W_{ext}    \label{eq_DeltaQ_WBL}
\end{align}
where $\bar{\alpha}=R\,[L]$ if $\alpha=L\,[R]$. Note that in Ref.\cite{Entin2017}, a similar expression was derived for the heat current in the peculiar case of a dot driven by a random telegraph noise $V(t)$ and after averaging over random processes. 
On the contrary, Eq.\eqref{eq_DeltaQ_WBL} is valid for an arbitrary $V(t)$ of finite support.
Finally, as it will be convenient in the following, we introduce the notation $w(E)=-\int_0^\infty\!\!\mathrm{d}t\,V(t)\partial_t|d(t,E)|^2$. We have $\int\!\mathrm{d}E\,w(E)=0$ and in virtue of Eq.\eqref{eq_P_general_WBL} 
\begin{equation}
\label{eq_Wext}
 W_{ext}=\int\!\frac{\mathrm{d}E}{2\pi}\left[\frac{f_L(E)}{\Gamma_R}+\frac{f_R(E)}{\Gamma_L}\right]w(E)\,.   
\end{equation}
Let us add a few comments. First, we check that 
\begin{equation}
\label{eq_1stlaw}
    \Delta Q_L+\Delta Q_R+(\mu_L-\mu_R)\Delta N_L+W_{ext}=0
\end{equation}
as expected from the first law of thermodynamics. 
Second, if the two reservoirs are at equilibrium (\textit{i.e.} $f_L=f_R$), then $\Delta N_L=\Delta N_R=0$ and $\Delta Q_\alpha$ reduces to  $\Delta Q_\alpha = \Gamma_\alpha / (\Gamma_\alpha + \Gamma_{\bar \alpha}) W_{ext}$. Thus, the input driving energy $W_{ext}$ is dissipated as heat in the two reservoirs\cite{moskalets2011}. The dissipation to the left and to the right reservoirs is asymmetric when the left-right symmetry is broken, here with $\Gamma_L\neq \Gamma_R$. On the contrary, if the two reservoirs are out-of-equilibrium (\textit{i.e.} $f_L\neq f_R$), the application of the pulse results in an additional transfer of particles ($\Delta N_L=-\Delta N_R\neq 0$) which is accompanied by a transfer of heat corresponding to the first term in the right-hand side of Eq.\eqref{eq_DeltaQ_WBL} (taking the form of a generalized Landauer formula). Since the external input energy $W_{ext}$ intervenes with a negative sign for both sides $L$ and $R$ in Eq.\eqref{eq_DeltaQ_WBL}, it can \textit{e.g} reduce the amount of heat leaving the hot reservoir and increase heat going to the cold reservoir (as we will see in Fig.\ref{fig_sqTrain_eff}).
Finally, to point out the role of an asymmetric coupling $\Gamma_L\neq \Gamma_R$ between the dot and the leads, it is convenient to introduce the parameters $a=(\Gamma_R-\Gamma_L)/(\Gamma_R+\Gamma_L)$, with $|a|\leq 1$, and $\Gamma=(\Gamma_L+\Gamma_R)/2$. With these notations,
\begin{equation}
    T_{dyn}(E,a)=(1-a^2)T_{dyn}(E,a=0)
\end{equation}
and
\mathleft
\begin{align}
\label{eq_Wext_alpha}
    W_{ext}(a)=\, & W_{ext}(a=0) \nonumber \\
            & -\frac{a}{\Gamma}\int\!\frac{dE}{2\pi}[f_L(E)\!-\!f_R(E)]w(E,a\!=\!0)\,.
\end{align}
\mathcenter
Thus $T_{dyn}$ is only re-normalized when the lead-dot coupling is made asymmetric while Eq.\eqref{eq_Wext_alpha} follows from energy conservation (\textit{i.e.} from Eq.\eqref{eq_1stlaw}). This leaves apparently no room for interesting effects of left/right asymmetry and our numerical exploration -- though empirical and non-exhaustive -- supports this statement. Therefore, except in Eqs.\eqref{eqs_d_squarepulse}, \eqref{eq_Tdyn(E)_square} \eqref{eq_w(E)_square} that we will keep general, we will assume in the rest of the manuscript and in all figures that left and right couplings are equal \textit{i.e.} $\Gamma_L=\Gamma_R=\Gamma$.

\begin{figure}[t]
    \centering
    \includegraphics[keepaspectratio,width=\columnwidth]{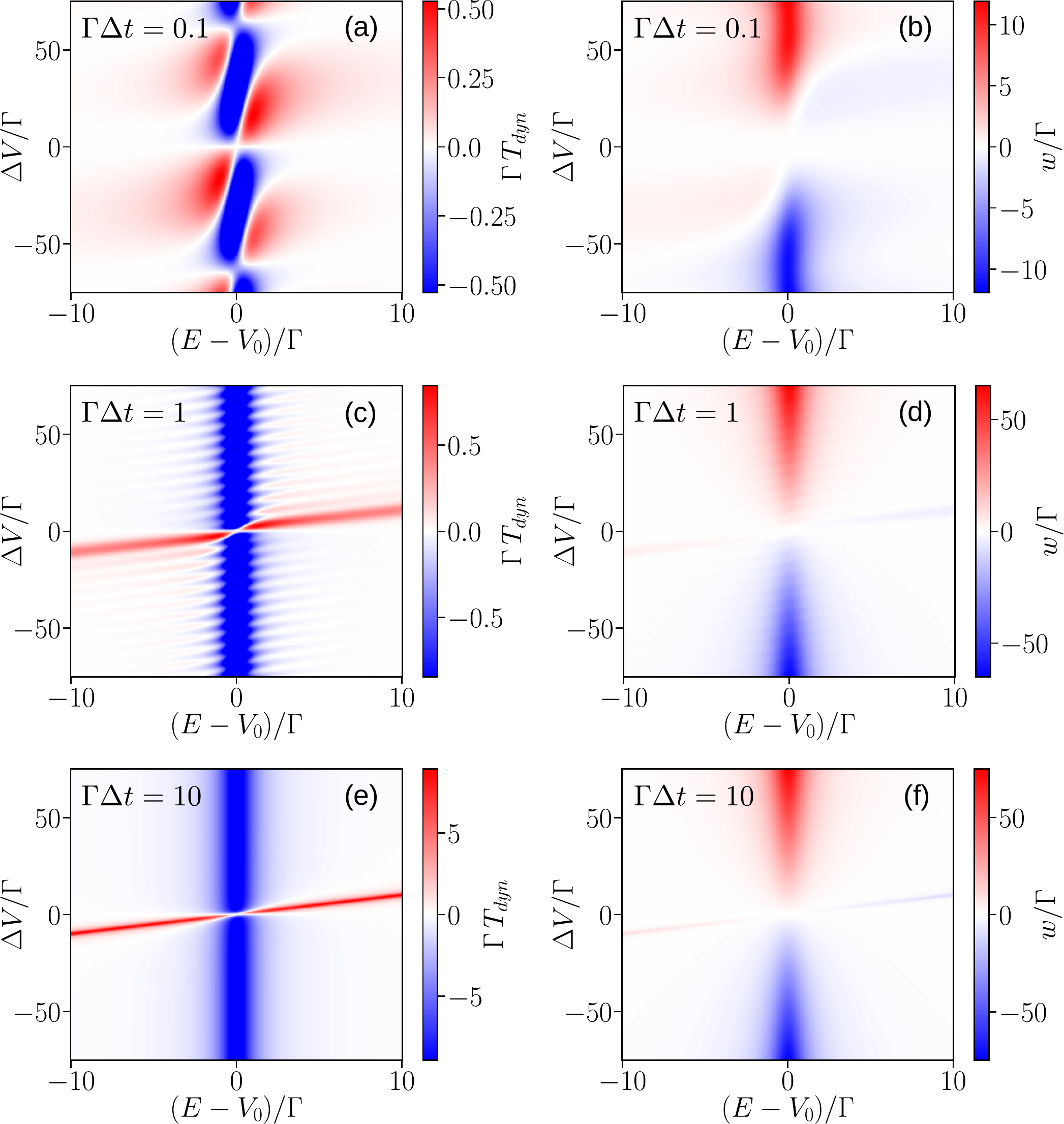}
    \caption{$T_{dyn}$ (left) and $w$ (right) for a square pulse (as given by Eqs.\eqref{eq_Tdyn(E)_square} and \eqref{eq_w(E)_square}), as a function of $E-V_0$ and $\Delta V$, for three values of $\Delta t$ ($0.1\Gamma^{-1}$ (top panels), $1\Gamma^{-1}$ (middle panels), and $10\Gamma^{-1}$ (bottom panels)). Taking $\Gamma$ as the energy unit yields colormaps that are independent of the value of $\Gamma$.}
    \label{fig_maps_Tdyn_w}
\end{figure}

\section{Single square pulse}
\label{sec_sq}
In this section, we assume that a single square pulse of amplitude $\Delta V$ and duration $\Delta t$ is applied on the dot, \textit{i.e.}
\begin{equation}
\label{eq_df_squarepulse}
    V(t) = \Delta V \Theta(t)\Theta(\Delta t -t)
\end{equation}
where $\Theta$ is the Heaviside function (with $\Theta(0)=1$). 
We push forward the semi-analytical approach described above and validate the results by comparing them to the ones obtained numerically with Tkwant. We study the convergence of the numerical results to the WBL and observe how the results are modified when the discontinuous jumps of the square pulse are made smooth.

\subsection{Semi-analytics in the wide-band limit}
\label{sec_analytics_sq}

When $V(t)$ is a square pulse, the transmission amplitude $d(t,E)$ takes a simple form in the WBL. It can be derived from Eq.\eqref{eq_d(t,E)} in Appendix \ref{appendixA} using \textit{e.g.} the residue theorem after having calculated the time integral defining $K$. We find (for arbitrary $\Gamma_L$, $\Gamma_R$)
\begin{align}
\label{eqs_d_squarepulse}
    d(t,E) =& ~ e^{-iEt}d_0(E)~~~~\mathrm{if}~t\leq 0, \nonumber \\
     d(t,E) =& ~e^{-iEt}d_1(E)+e^{-(\Gamma+iV_1) t}d_{01}(E) \nonumber \\
           &  \hspace{4cm}\mathrm{if}~0\leq t\leq \Delta t,   \nonumber \\
     d(t,E) =& ~e^{-iEt}d_0(E)+\left\{[e^{-(\Gamma+iV_1) \Delta t}-e^{-i\Delta t E}]\times\right. \nonumber \\
           &   \left.e^{-[\Gamma+iV_0](t-\Delta t)}d_{01}(E)\right\}~~~~\mathrm{if}~\Delta t \leq t\,.
\end{align}
Here we have introduced the notations $V_1=V_0+\Delta V$, $d_1(E)=d_0(E-\Delta V)$, and $d_{01}(E)=d_0(E)-d_1(E)$, and as before $\Gamma=(\Gamma_L+\Gamma_R)/2$. $T_{dyn}(E)$ and $w(E)$ introduced in Sec.\ref{sec_timeintegrated_WBL} can be calculated analytically as well. Using Eqs.\eqref{eq_d0(E)} and \eqref{eqs_d_squarepulse}, it can be shown that
\begin{widetext}
\mathleft
\begin{align}
    T_{dyn}(E) & = D_{10}\Delta t+\left[e^{-\Gamma\Delta t}\cos(\beta)\!-\!1\right]\left[\frac{2\Gamma}{\Gamma_L\Gamma_R}D_{10}^2-\!\frac{\Delta V^2}{\Gamma}\frac{D_0D_1}{\Gamma_L\Gamma_R}\right]
      -2e^{-\Gamma\Delta t}\sin(\beta)D_{10}\left[\frac{E\!-\!V_0}{\Gamma_L\Gamma_R}D_0+\frac{V_1\!-\!E}{\Gamma_L\Gamma_R}D_1\right] \label{eq_Tdyn(E)_square} \\
    w(E) & = -D_{10}\Delta V+\frac{\Delta V^2}{\Gamma_L\Gamma_R}D_0D_1  \left[
    2e^{-\Gamma\Delta t}\left[(E-V_0)\cos(\beta)-\Gamma\sin(\beta)\right]-\Delta V e^{-2\Gamma\Delta t}\right]  \label{eq_w(E)_square}
\end{align}
\mathcenter
\end{widetext}
where $D_0(E)=|d_0(E)|^2$, $D_1(E)=|d_1(E)|^2$, $D_{10}(E)=D_1(E)-D_0(E)$, and $\beta(E)=\Delta t(V_1-E)$. The two quantities are plotted in Fig.\ref{fig_maps_Tdyn_w} as a function of $E-V_0$ and $\Delta V$ for three values of $\Delta t$. They can be positive or negative and satisfy the sum rules $\int\!\mathrm{d}E\,T_{dyn}(E)=\int\!\mathrm{d}E\,w(E)=0$ as noticed in Sec.\ref{sec_timeintegrated_WBL}. For large $\Delta t \gg \Gamma^{-1}$ (bottom panels in Fig.\ref{fig_maps_Tdyn_w}),  $T_{dyn}$ and $w$ reduce formally to their adiabatic limits, \textit{i.e.} $T_{dyn}\approx D_{10}\Delta t$ and $w\approx -D_{10}\Delta V$ (though the variation of the dot energy level at $t=0$ and $\Delta t$ occurs instantaneously). When $\Delta t$ becomes comparable to $\Gamma^{-1}$ or smaller, the colormaps of $T_{dyn}$ and $w$ get modified and in particular, oscillations appear. 
Besides, a thorough graphical analysis of the function $w(E)$ allows us to conclude that $\int\!\mathrm{d}E\,f_\alpha(E)w(E)\geq 0$ (knowing that $\int\!\mathrm{d}E\,w(E)=0$), and hence that $W_{ext}\geq 0$ for a square pulse in virtue of Eq.\eqref{eq_Wext}. Thus, the source driving the dot supplies energy to the junction, which eventually heats up the reservoirs.

\subsection{Analytics vs numerics}
\label{sec_analyticsvsnum}

\begin{figure}[t]
    \centering
    \includegraphics[keepaspectratio,width=\columnwidth]{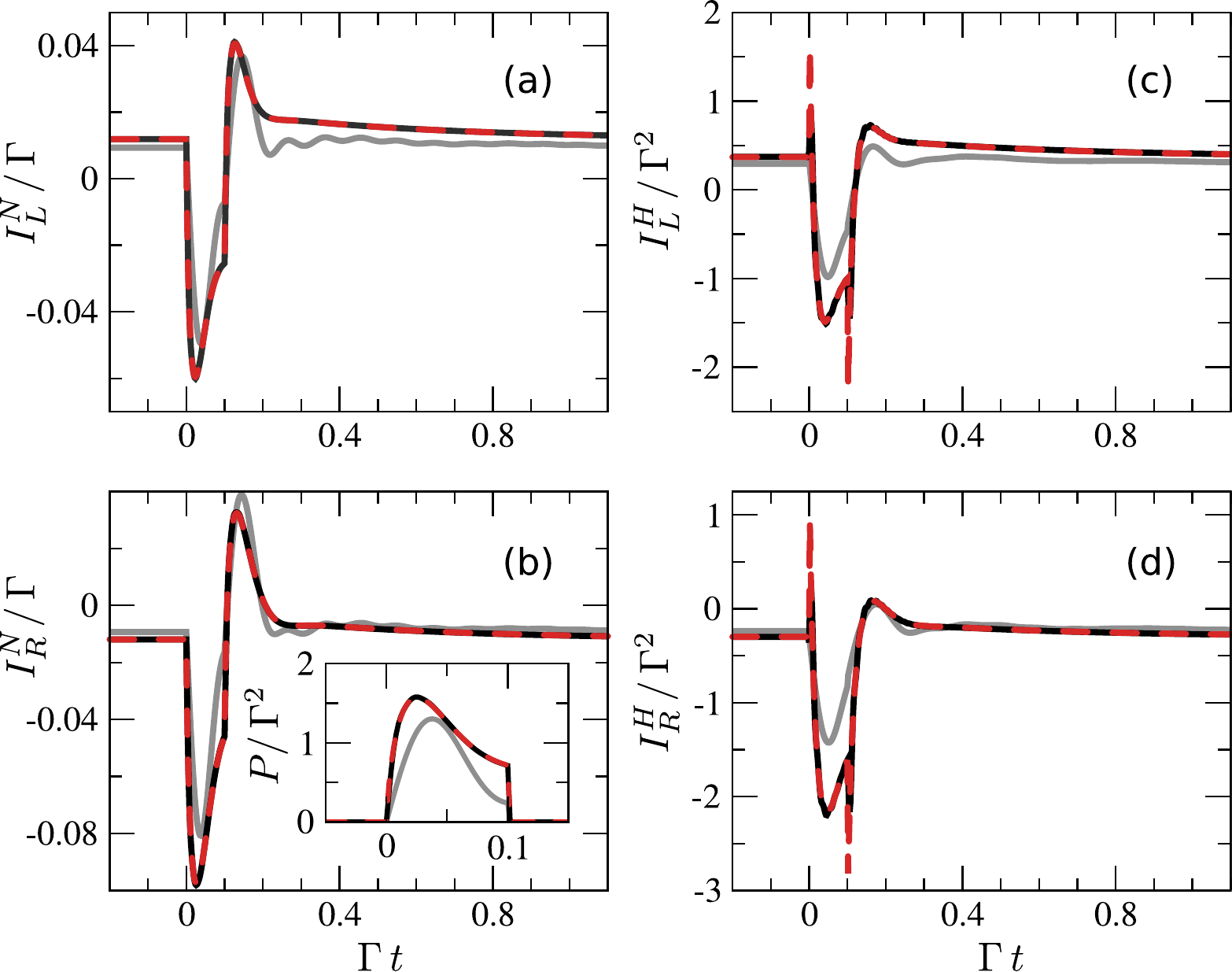}
    \caption{$I^N_L$ (a), $I^N_R$ (b), $I^H_L$ (c), $I^H_R$ (d), and $P$ (inset in (b)), as a function of time $t$, when a single square pulse is applied in the dot. In all panels, numerical data computed with Tkwant for two values of $\lambda$ ($20$ (gray lines) 
    and $200$ (black lines)) are compared to the semi-analytical results (red dashed lines) valid in the WBL $\lambda\to\infty$ and given by Eqs.\eqref{eq_IN(t)_1}-\eqref{eq_P_general_WBL}. Each line is built up from evenly spaced points with a time step $\delta t=0.002\Gamma^{-1}$. With the exception of the heat current plots around singularities at $t=0$ and $t=\Delta t$ (see Fig.\ref{fig_IHLcusp}(a) for a zoom), the red and black lines are superimposed. Parameters: $\Gamma_L=\Gamma_R=\Gamma$,  $T_L=12\Gamma$, $T_R=8\Gamma$, $\mu_L=-3\Gamma$, $\mu_R=3\Gamma$, $V_0=25\Gamma$, $\Delta V=10\Gamma$, $\Gamma\Delta t=0.1$, and $\gamma=\Gamma$.}
    \label{fig_all_vs_t}
\end{figure}

\begin{figure}[t]
    \centering
    \includegraphics[keepaspectratio,width=\columnwidth]{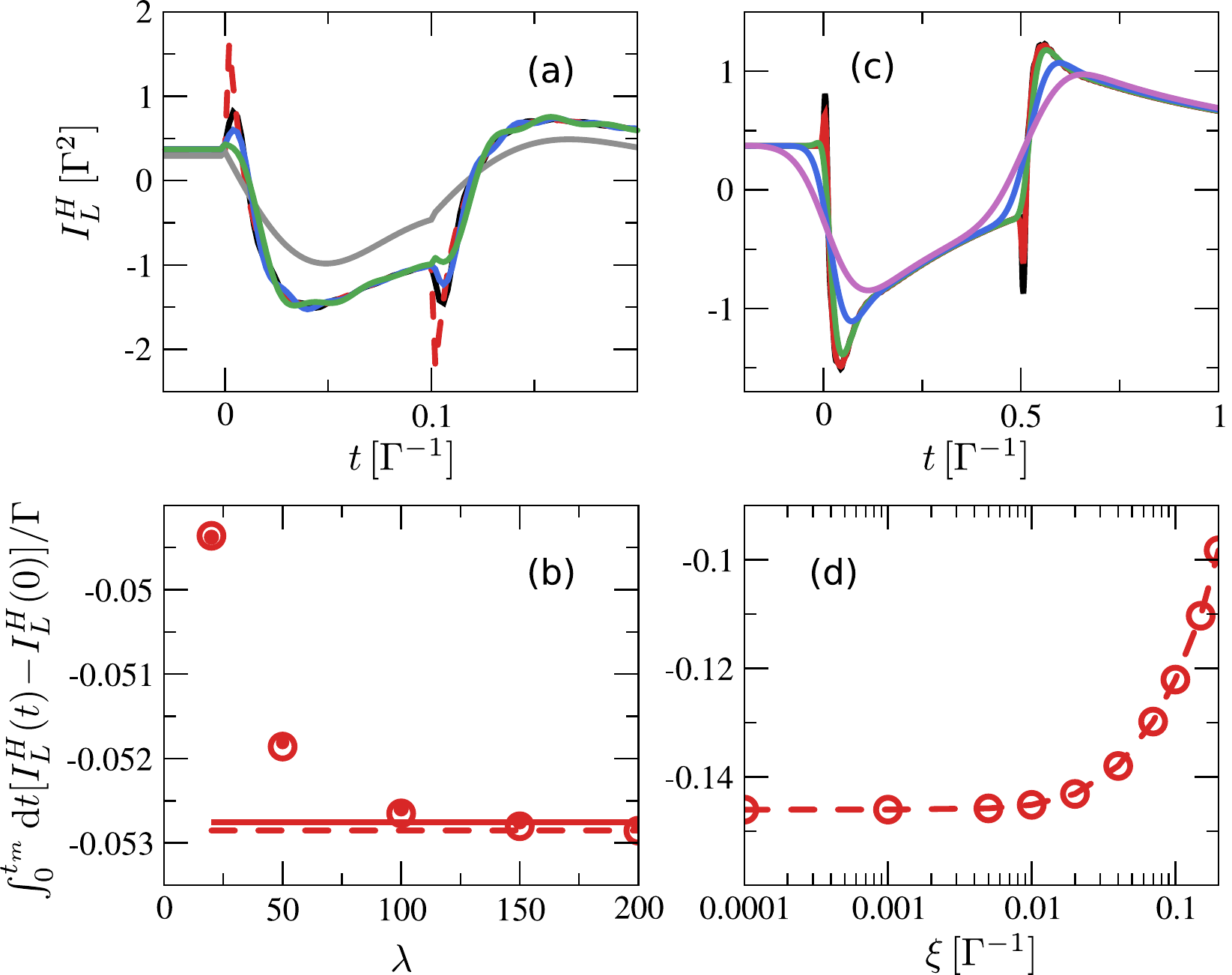}
    \caption{
    (a) Zoom of Fig.\ref{fig_all_vs_t}(c) with additional Tkwant data for $\lambda=100$ (green line) and $\lambda=150$ (blue line). The amplitude of the cusps at $t=0$ and $t=\Delta t$ increases with $\lambda$. (b) Integral $\int_0^{t_m}\mathrm{d}t[I^H_L(t)-I^H_L(0)]/\Gamma$ of data shown in (a) as a function of $\lambda$, with $t_m=1.75\Gamma^{-1}$. Empty [full] circles correspond to the numerical (Simpson's) integration of Tkwant data with a time step $\delta t=0.002\Gamma^{-1}$ [$\delta t=0.0001\Gamma^{-1}$]. The dashed line indicates the value of the integral calculated in the WBL with Simpson's rule using points of the dashed line in (a) separated by $\delta t=0.002\Gamma^{-1}$. The full line indicates this value when the integral over time is done analytically. (c) $I^H_L(t)$ calculated with Tkwant for a smoothed square pulse, with different values of the characteristic length $\xi$ ($\Gamma\xi=0.0001$ (black line), $0.01$ (red line), $0.04$ (green line), $0.1$ (blue line), and $0.2$ (purple line)) and for $\Gamma\Delta t=0.5$ and $\lambda=200$. Other parameters are the same as in Fig.\ref{fig_all_vs_t}. (d) Integral $\int_0^{t_m}\mathrm{d}t[I^H_L(t)-I^H_L(0)]/\Gamma$ of data shown in (c) as a function of $\xi$ (circles) with $t_m=2.5\Gamma^{-1}$ (for $t\geq 2.5\Gamma^{-1}$, $I^H_L(t)$ is (almost) independent of $\xi$). The dashed line serves as a guide to the eye.}
    \label{fig_IHLcusp}
\end{figure}

\begin{figure}[t]
    \centering
    \includegraphics[keepaspectratio,width=\columnwidth]{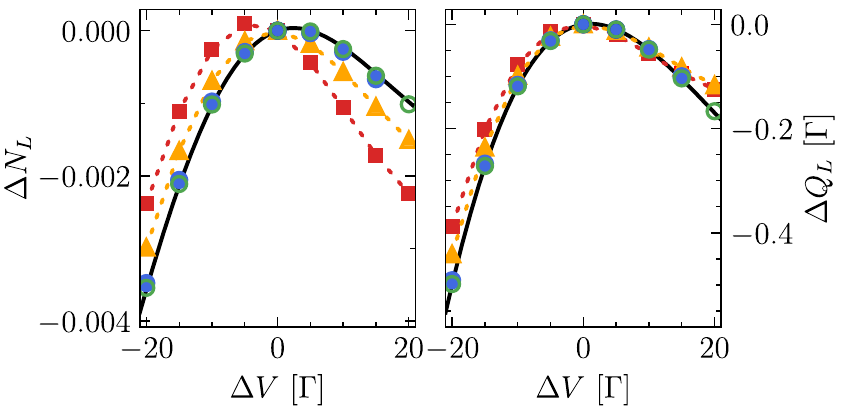}
    \caption{
    $\Delta N_L$ (left) and $\Delta Q_L$ (right) as a function of the square pulse amplitude $\Delta V$, others parameters being the same as in Figs.\ref{fig_all_vs_t} and \ref{fig_IHLcusp}. The black lines show the analytical predictions in the WBL given by Eqs.\eqref{eq_DeltaN}-\eqref{eq_Wext}, \eqref{eq_Tdyn(E)_square},  and \eqref{eq_w(E)_square} while the symbols correspond to Tkwant data for $\lambda=15$ (red squares), $20$ (orange triangles), $50$ (blue dots), and $100$ (green circles). Dotted lines are guides to the eye.
    } 
    \label{fig_anaVsnum_int}
\end{figure}

By inserting Eq.\eqref{eqs_d_squarepulse} into Eqs.\eqref{eq_IN(t)_1}-\eqref{eq_P_general_WBL}, we can compute semi-analytically $I^N_\alpha(t)$, $I^H_\alpha(t)$, and $P(t)$, the integrals over the energy being done numerically. The results are plotted in Fig.\ref{fig_all_vs_t} for a given configuration and are compared to the exact Tkwant results for two values of $\lambda$. 
Note that the analytical results in the WBL are independent of $\Gamma$ when energies and times are expressed in units of $\Gamma$ and $\Gamma^{-1}$ respectively, while the Tkwant results do not depend on $\lambda$, $\gamma$, and $\Gamma$ separately but only on the ratio $\lambda\gamma/\Gamma$. In the limit of large $\lambda(=200)$ corresponding to the WBL, the two sets of curves overlap perfectly, with the exception of the heat currents near $t=0$ and $t=\Delta t$. Indeed in the WBL, the heat currents display singularities\cite{covito2018,kara2020} at those two times where $V(t)$ jumps instantaneously from $0$ to $\Delta V$ and conversely. This is highlighted in Fig.\ref{fig_IHLcusp}(a) where data are zoomed in. However, it can be shown analytically that those two singularities are regularized after integration over time \textit{i.e.} $\int_{-\infty}^{t_m}\!\mathrm{d}t\,[I^H_\alpha(t)-I^H_\alpha(t=0)]$ is well defined. As illustrated in Fig.\ref{fig_IHLcusp}(b), the $\lambda-$dependence of the Tkwant curves $I^H_\alpha(t)$ near the singularities becomes irrelevant in the limit of large $\lambda$ once the heat currents are integrated over time and very good agreement is found between analytics and numerics in the WBL after integration.\\
\indent In Fig.\ref{fig_IHLcusp}(c), we also study with Tkwant how the time-resolved heat current $I^H_L(t)$ is modified when $V(t)$ is not varied abruptly but continuously between $0$ and $\Delta V$, as
\begin{equation}
\label{eq_df_smoothsquarepulse}
    V(t) = \frac{\Delta V}{2}\left[\mathrm{erf}\left(\frac{t}{\xi/2}\right)-\mathrm{erf}\left(\frac{t-\Delta t}{\xi/2}\right)\right]
\end{equation}
$\mathrm{erf}$ being the error function and $\xi$ ($\lesssim \Delta t$) the characteristic time controlling the smoothness of the square-like pulse. Data are plotted for $\lambda=200$ (\textit{i.e} near the WBL) and different values of $\xi$. We find that $I^H_L(t)$ is (almost) independent of $\xi$, as long as $\xi\lesssim 0.01\Gamma^{-1}$. Small $\xi$-dependence is only visible in that range near the cusps at $t=0$ and $t=\Delta t$ and their amplitude turns out to decrease with $\xi$. As shown in Fig.\ref{fig_IHLcusp}(d), after integration over time, $\int I^H_L(t)\mathrm{d}t$ is strictly independent of $\xi$ for $\xi\lesssim 0.01\Gamma^{-1}$. This is also true for the other quantities $I^H_R(t)$, $I^N_L(t)$, $I^N_R(t)$, and $P(t)$ (data not shown). Let us briefly discuss what this implies for experiments. Realistic square pulses that can be applied in experiments takes the form of Eq.\eqref{eq_df_smoothsquarepulse} while the theoretical limit $\xi\to 0$ yielding the square pulse \eqref{eq_df_squarepulse} cannot be engineered. Taking\cite{Josefsson2018} $\Gamma\approx 10$\,GHz for a realistic quantum dot operating at an average temperature $T=10\Gamma\approx 0.5$\,K (to be consistent with parameter values taken in Fig.\ref{fig_IHLcusp}), we find that the square pulse model is relevant for discussing such experiments if $\xi\lesssim 1$ps which is within reach (yet challenging) experimentally\cite{Bauerle_2018}.\\
\indent Finally, we have also validated analytical formulas \eqref{eq_DeltaN}-\eqref{eq_Wext}, \eqref{eq_Tdyn(E)_square},  and \eqref{eq_w(E)_square} for $\Delta N_\alpha$, $\Delta Q_\alpha$, and $W_{ext}$ with the help of Tkwant simulations. This is illustrated in Fig.\ref{fig_anaVsnum_int} where $\Delta N_L$ and $\Delta Q_L$ are plotted as a function of the square pulse amplitude $\Delta V$, other parameters being fixed. We observe that the convergence of Tkwant data to the analytical curves valid in the WBL depend on $\Delta V$. Indeed, the square driving makes transport inelastic within an energy range centered around $\mu_L\approx\mu_R$ of width $\sim\!\Delta V$. When $\Delta V$ increases, some outgoing plane waves are excited at energies $E'$ closer and closer to the boundaries $\pm 2\gamma\lambda$ of the lead conduction band and $\lambda$ must be increased as well to reach the WBL.

\section{Heat engine efficiency}
\label{sec_eff}

\begin{figure}[t]
    \centering
    \includegraphics[keepaspectratio,width=\columnwidth]{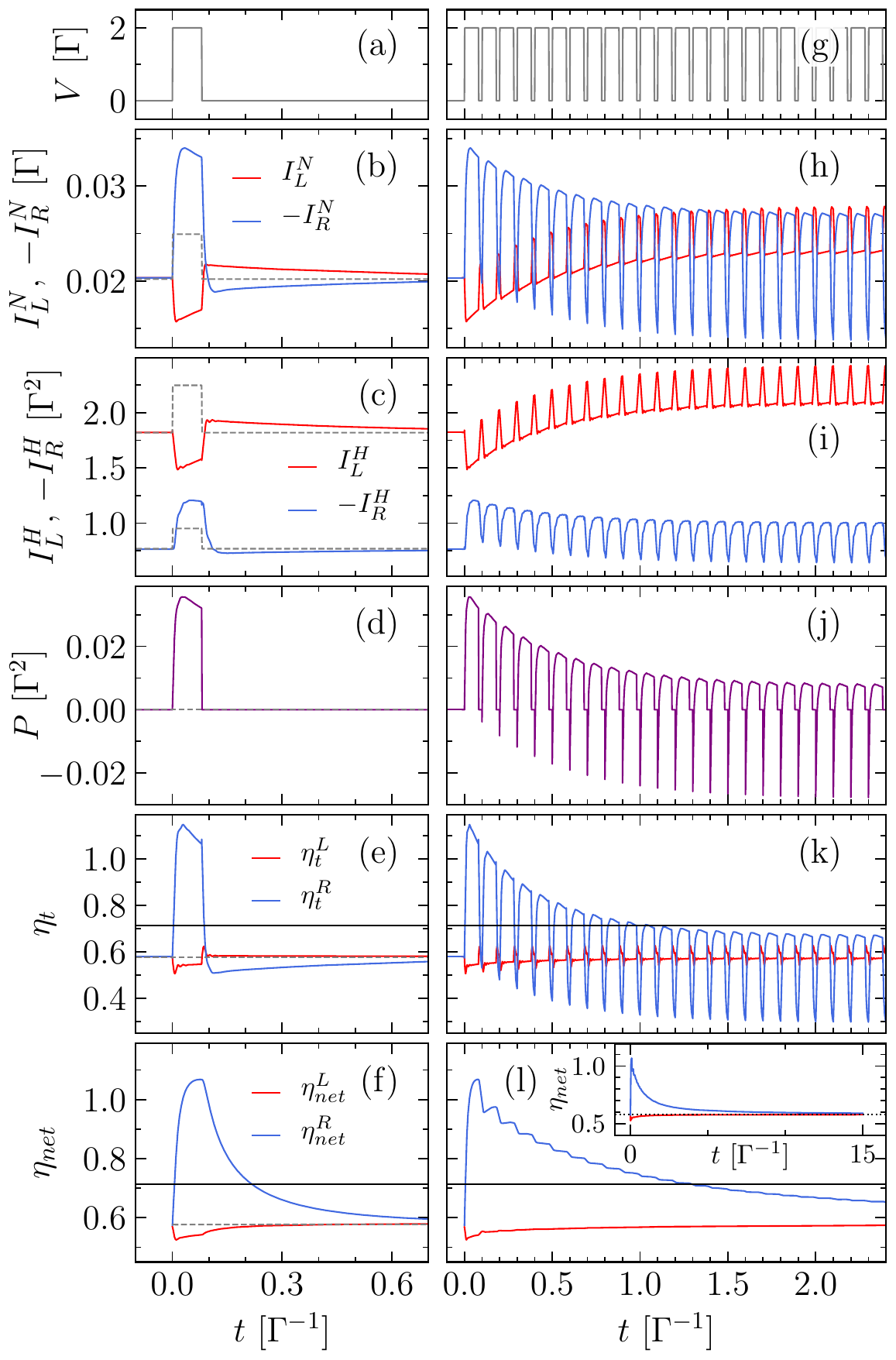}
    \caption{Transient increase of the heat engine efficiency. (Left) Particle currents $I^N_{L,R}$ (b), heat currents $I^H_{L,R}$ (c), input power $P$ (d), instantaneous efficiency $\eta_t$ (e), and net efficiency $\eta_{net}$ (f) as a function of time $t$ when the single square pulse $V(t)$ shown in (a) is applied. The different curves in (e) and (f) correspond to different definitions of the efficiencies ($\eta_{t/net}^L$ (in red), $\eta_{t/net}^R$ (in blue), see main text). The horizontal black line indicates the Carnot efficiency value $\eta_C$. In panels (b) to (f), each quantity is also evaluated within the quasi-static approximation in the WBL (gray dashed lines). Parameters: $\Gamma_L=\Gamma_R=\Gamma$, $T_L=87\Gamma$, $T_R=25\Gamma$, $\mu_L=-26\Gamma$, $\mu_R=26\Gamma$, $V_0=55\Gamma$, $\Delta V=2\Gamma$, $\Delta t=0.08/\Gamma$, $\gamma=\Gamma$, and $\lambda=200$. (Right) Same as left panels when the square pulse $V(t)$ is repeated periodically for $t>0$, with a period $\tau=0.1/\Gamma$, as shown in (g). The inset in (l) shows the convergence of $\eta_{net}$ to the steady state efficiency (black dotted line). } 
    \label{fig_sqTrain_eff}
\end{figure}

In this section, we focus on standard Seebeck configurations with \textit{e.g.} $T_L>T_R$ and $\mu_L<\mu_R$, and investigate whether driving the dot energy level with $V(t)$  in the time-dependent RLM is beneficial or detrimental to the heat engine efficiency. 
We restrict our study to the WBL for which semi-analytical formula have been derived above. We consider in Fig.\ref{fig_sqTrain_eff} a set of parameters $T_L>T_R$, $\mu_L<\mu_R$, and $V_0$ such that the system in the WBL is in a Seebeck configuration for $t<0$, with a high heat-to-work (stationary) thermoelectric efficiency $\eta_{st}=\Delta\mu I^N_L(t<0)/I^H_L(t<0)\approx 0.58 \approx 0.81\eta_C$, where $\Delta\mu\equiv \mu_R-\mu_L$ and $\eta_C=1-T_R/T_L$ is the Carnot efficiency. Then for $t>0$, we switch on the time-dependent perturbation $V(t)$ and consider first that $V(t)$ is a single square pulse as shown in Fig.\ref{fig_sqTrain_eff}(a). Here the amplitude $\Delta V$ of the square pulse is chosen small enough so that in an adiabatic picture, the stationary RLM with a dot energy level $\epsilon_0=V_0+\Delta V$ remains in a Seebeck configuration. Moreover its heat-to-work efficiency remains unchanged ($\pm 0.001$) with respect to the stationary configuration with $\epsilon_0=V_0$ at $t<0$. However, we will see hereafter that the exact transient response of the RLM (beyond the adiabatic limit) is significantly impacted by the time-dependent driving. In panels (b), (c), and (d) of Fig.\ref{fig_sqTrain_eff}, we plot with full lines $I^N_\alpha(t)$, $I^H_\alpha(t)$, and $P(t)$ computed with Tkwant taking $\lambda=200$. Data are in very good agreement with the ones (not shown) obtained from Eqs.\eqref{eq_IN(t)_1}-\eqref{eq_P_general_WBL} and \eqref{eqs_d_squarepulse}. Additionally, we plot with dashed lines the corresponding quantities $I^{N,ad}_\alpha(t)$, $I^{H,ad}_\alpha(t)$, and $P^{ad}(t)$ in the adiabatic (quasi-static) approximation. Obviously, $P^{ad}(t)=0$ while the adiabatic particle and heat currents are given by the stationary Landauer-B\"uttiker formula
\begin{align}
    I^{N,ad}_\alpha(t) & = \int\!\frac{\mathrm{d}E}{2\pi}D(t,E)[f_\alpha(E)-f_{\bar{\alpha}}(E)] \\
    I^{H,ad}_\alpha(t) & = \int\!\frac{\mathrm{d}E}{2\pi}(E-\mu_\alpha)D(t,E)[f_\alpha(E)-f_{\bar{\alpha}}(E)]
\end{align}
($\bar{\alpha}=R\,[L]$ if $\alpha=L\,[R]$) with a transmission probability $D(t,E)$ depending parametrically on time: $D(t,E)=D_1(E)$ for times $0\leq t\leq \Delta t$ and $D(t,E)=D_0(E)$ elsewhere, $D_0(E)$ and $D_1(E)$ being defined in Sec.\ref{sec_analytics_sq}. The Tkwant curves are qualitatively different from the adiabatic ones. When $V(t)$ is increased from $0$ to $\Delta V>0$, electrons of higher energy are expelled from the dot to both baths which leads to a decrease of $I^{N}_L(t)$ and $I^{N}_R(t)$ (and conversely when $V(t)$ goes back to $0$). The same effect is observed for the heat currents $I^{H}_L(t)$ and $I^{H}_R(t)$. This peculiar behavior due to displacement particle and energy currents renders difficult a proper definition of an efficiency, as we will see in the following. Besides, $0\leq P(t)\ll I^{H}_L(t), |I^{H}_R(t)|$ for all times.

We now tackle the question of the thermodynamic efficiency of this driven heat engine. Defining such an efficiency in the time-dependent regime is particularly subtle. Here we use and comment on different possible definitions. We introduce first an instantaneous (time-resolved) efficiency $\eta_t^{R}(t)=\Delta\mu(-I^N_R(t))/(I^H_L(t)+P(t))$ plotted in blue in Fig.\ref{fig_sqTrain_eff}(e). Its denominator is the sum of the resources used to generate electric power, \textit{i.e.} the input power $P(t)\geq 0$ and the heat current $I^H_L(t)>0$ leaving the hot bath. Importantly, for this parameter set, $I^H_L(t)>0$ and $I^H_R(t)<0$ for all times. 
Its numerator is the electric power generated by electrons flowing out towards the (right) bath of higher electrochemical potential $\mu_R>\mu_L$. For completeness, we also plot in red in Fig.\ref{fig_sqTrain_eff}(e) the instantaneous efficiency $\eta_t^{L}(t)=\Delta\mu I^N_L(t)/(I^H_L(t)+P(t))$.  
While $\eta_t^L(t)$ is only slightly affected by the driving, $\eta_t^R(t)$ increases drastically in the transient regime, above $\eta_C$ and even above $1$. Actually, we argue that such instantaneous (time-resolved) efficiencies are ill-defined for two main reasons. First, one may dispute the choice of the numerator in $\eta_t^L(t)$ or $\eta_t^R(t)$. In a stationary Seebeck configuration, the generated electric power is attributed to electrons leaving the hot bath ($T_L>T_R$) and climbing an electrochemical potential bias $\Delta\mu=\mu_R-\mu_L>0$. In the (non-interacting) time-dependent case, this picture is modified due to the presence of the displacement current (\textit{i.e.} temporary particle storage in the dot, see Eq.\eqref{eq_chargecsv}) and one should include the electromagnetic environment of the device to define properly the output power of the driven heat engine. Second, the definition of the time-resolved heat current $I^H_L(t)$ intervening in the denominator of $\eta_t^L(t)$ or $\eta_t^R(t)$ is also controverted. In \textit{e.g.} Refs.\cite{crepieux2011,zhou2015}, the last term in the right hand side of Eq.\eqref{eq_IH(t)_WBL} due to the lead-dot coupling contribution is not included\footnote{While $I^H_L(t)$ defined by Eq.\eqref{eq_df_IH(t)} is given by Eq.\eqref{eq_IH(t)_WBL} in the WBL, the alternative heat current $\tilde{I}^H_\alpha(t)=-\frac{\mathrm{d}}{\mathrm{d}t}\langle H_\alpha\rangle-\mu_\alpha I^N_\alpha(t)$ is given by the same equation without the last term in its right hand side. Our simulations for a square pulse $V(t)$ show that both quantities differ near $t=0$ and (oppositely) near $t=\Delta t$ when $V(t)$ varies abruptly. When time is integrated out, the two behaviours cancel each other and we check the sum rule, $\int\! \mathrm{d}t\, I^H_\alpha(t)=\int\! \mathrm{d}t\, \tilde{I}^H_\alpha(t)$. The two heat currents definitions lead to two instantaneous efficiencies $\eta_t^\alpha(t)$ ($\alpha=L$ or $R$) that differ also near $t=0$ and $t=\Delta t$. Those discrepancies -- that are local in time and cancel out at long times -- are suppressed in the net efficiencies $\eta_{net}^\alpha(t)$.}. Besides, $I^H_L(t)$ is defined at the dot-lead interface while heat is eventually dissipated later on and on a different timescale in the bath attached to it\cite{Oz2020}. Moreover, one may argue that only the thermoelectric contribution to $I^H_L(t)$ (\textit{i.e.} the first term in the right hand side of Eq.\eqref{eq_IH(t)_WBL}) and not the whole $I^H_L(t)$ should be considered as a resource for the engine since the driving -- whose contribution is counted by adding $P(t)$ in the denominator of $\eta_t^L(t)$ or $\eta_t^R(t)$ -- also heats up the baths and hence contributes to $I^H_L(t)$. Before ending this discussion, we put forward an alternative definition of the efficiency by noticing that the relevant quantity for application purposes is not the instantaneous efficiency but the net efficiency of the engine at the end of the experiment. Thus we introduce
\begin{align}
    N_\alpha(t) &=\int_0^{t}\!\! \mathrm{d}u\,I^N_\alpha(u) \\
    Q_\alpha(t) &=\int_0^{t}\!\! \mathrm{d}u\,I^H_\alpha(u) \\
    W_{ext}(t) &=\int_0^{t}\!\! \mathrm{d}u\,P(u) 
\end{align}
which correspond to the net number of particles ($N_\alpha(t)$) and the net heat ($Q_\alpha(t)$) flowing out from lead $\alpha$, as well as the net driving work ($W_{ext}(t)$), measured from time $t=0$ at which $V(t)$ is switched on to the running time $t$. Then we define the two corresponding net efficiencies $\eta_{net}^L(t)=\Delta\mu N_L(t)/(Q_L(t)+W_{ext}(t))$ and $\eta_{net}^R(t)=\Delta\mu (-N_R(t))/(Q_L(t)+W_{ext}(t))$ plotted respectively in red and in blue in Fig.\ref{fig_sqTrain_eff}(f).
Note that $W_{ext}(t)\ll Q_L(t)$ so that its contribution in the denominator of $\eta_{net}^\alpha(t)$ and in $Q_L(t)$ (given at long times by Eq.\eqref{eq_DeltaQ_WBL}) is negligible. We find that $\eta_{net}^L(t)$ is always (slightly) smaller than the stationary efficiency $\eta_{st}$. On the contrary, $\eta_{net}^R(t)$ increases drastically in the transient regime. The behavior of these two efficiencies can be qualitatively interpreted: when the dot level undergoes the Heaviside jump upwards, the now higher-energy electrons residing in it temporarily flow outwards to both leads. This has the transient effect of \textit{(i)} increasing the net number of electrons (climbing the chemical potential step) that flow to the cold lead ; \textit{(ii)} decreasing the net particle current leaving the hot lead ; \textit{(iii)} increasing the net heat current going to the cold lead ; \textit{(iv)} decreasing the net heat leaving the hot lead. These explain both the behavior of $\eta_{net}^R(t)$ and $\eta_{net}^L(t)$ during the upwards jump, a similar analysis can be made for the downward jump that follows. Finally, both $\eta_{net}^L(t)$ and $\eta_{net}^R(t)$ converge to $\eta_{st}$ at long times.\\ 
\indent Since $\eta_{net}^R(t)$ can be increased at short times by driving the dot with a single square pulse, let us now investigate the same device when $V(t)$ is cycled as 
\begin{equation}
\label{eq_df_squaretrain}
    V(t) = \Delta V \sum_{n\in\,\mathbb{N}} \Theta(t-n\tau)\Theta(n\tau+\Delta t -t)\,.
\end{equation}
The period $\tau$ of the pulse train is chosen so as $\eta_{net}^R(t=\tau)$ in Fig.\ref{fig_sqTrain_eff}(f) is close to its maximum. Our results are summarized in the right panels of Fig.\ref{fig_sqTrain_eff}. We see in particular in panel (l) that after the transient increase of $\eta_{net}^R(t)$ at short times, the convergence time of $\eta_{net}^R(t)$ to the steady-state efficiency $\eta_{ss}=\Delta\mu N_L^{\tau,\infty}/(Q_L^{\tau,\infty}+W_{ext}^{\tau,\infty})$ is much larger than the one of $\eta_{net}^R(t)$ to $\eta_{st}$ in the case of a single square pulse (see panel (f)). However, we find $\eta_{ss}\approx \eta_{st}$ (with a discrepancy of about $0.002$). Thus the transient increase of the efficiency which may appear advantageous for thermoelectric applications is eventually cancelled out in the long time limit. Roughly speaking, what is won at the beginning is eventually given back. 

We end up this section with a remark concerning the validity range of our results. We focused the discussion above on the analysis of Fig.\ref{fig_sqTrain_eff} valid for a given set of parameters. Yet actually, we also performed a systematic study in the case of a single square pulse by deriving analytical expressions of $N_\alpha(t)$, $Q_\alpha(t)$, and $W_{ext}(t)$ in the WBL. Though such expressions are lengthy (not shown), their numerical evaluation is almost immediate whereas using Tkwant with a large $\lambda$ and integrating subsequently over time is much more expensive in computation time. This approach allowed us to investigate about 200000 configurations in the parameter space $(T_L>T_R, \mu_L<\mu_R, V_0, \Delta V, \Delta t)$, the choice of $\Gamma$ value playing no role after proper scaling as noticed in Sec.\ref{sec_analyticsvsnum}. We restricted ourselves to parameter sets leading to Seebeck configurations in the stationary cases  $\epsilon_0=V_0$ and $\epsilon_0=V_0+\Delta V$, and explored randomly the parameter space by keeping roughly equidistant points with respect to the corresponding stationary efficiencies. We used for that purpose the Adaptive Python package\cite{Nijholt2019}. We monitored $\eta_{net}^R(t)$ and found many configurations yielding behaviors qualitatively similar to the one shown in Fig.\ref{fig_sqTrain_eff}(f), while other configurations showed a transient decrease of $\eta_{net}^R(t)$ or oscillating decaying behaviors around $\eta_{ss}$. The parameter set in Fig.\ref{fig_sqTrain_eff} was chosen randomly among the most promising ones \textit{i.e.} the ones giving the largest transient net efficiency and the largest increase with respect to $\eta_{ss}$, keeping $\eta_{net}^R(t)>\eta_{ss}$ for all times. Then, only this configuration was subjected to a cyclic square driving. Moreover, the value of the period $\tau$ was chosen so as to increase significantly the convergence time of $\eta_{net}^R$ to $\eta_{ss}\approx\eta_{st}$. Other simulations were also run for a few other values of $\tau$ and showed similar behaviours \textit{i.e.} a convergence of the net efficiency to a constant value very close to $\eta_{st}$. Hence, as discussed above, we found no advantage on the net efficiency of driving the dot as the transient increase of $\eta_{net}^R(t)$ cannot be leveraged in practical heat engines that need to operate for long times.
Our conclusion is nevertheless purely empirical and still lacks a rigorous proof.

\section{Conclusion}
\label{sec_ccl}

We have studied thermoelectric transport in the non-interacting time-dependent RLM. We have first used our scattering approach to construct a compact analytical framework describing thermoelectric transport within the WBL approximation but beyond the adiabatic and weak coupling limits. Our time-resolved expressions of particle and heat currents as well as input power are the counterparts in the scattering approach of formulas previously derived in the literature\cite{jauho1994,crepieux2011,zhou2015,dare2016,kara2020} using the non-equilibrium Green's function technique. Our time-integrated formulas also reproduce formally expressions given in previous works\cite{gaury2014a,Gurvitz2019,Entin2017} though the energy-related quantities were calculated in Ref.\cite{Entin2017} for a telegraph noise while Eqs.\eqref{eq_DeltaQ_WBL} and \eqref{eq_Wext} are valid for arbitrary pulses (of finite duration).  Moreover, we have pushed forward our analytical approach in the peculiar case where the dot is driven with a single square pulse and performed numerical simulations to study the convergence towards the WBL and deviations appearing when the square pulse is smoothed. 
Our analytics allowed us to spot interesting regimes that we studied numerically in a second step so as to address the cases where the pulse is repeated periodically. 
Considering a device in a stationary Seebeck configuration, we have studied the possibility of enhancing the efficiency of the heat engine by time-dependent driving. We have observed that the net thermodynamic efficiency of the device, defined over a cycle of a periodic drive or over the duration of an experiment (\textit{i.e.} after switching on and before switching off the driving), is not enhanced with respect to the stationary efficiency. The advantage brought by the driving that may appear in the transient regime at short times is lost 
in the long run. It is noteworthy that, at a rough qualitative level, similar conclusions were drawn in Ref.\cite{Snyder2002} reporting on classical experiments of Peltier cooling driven by a current pulse.\\ \indent Our work contributes to the growing literature using the RLM as a test bed to investigate transport and thermodynamics in out-of-equilibrium nanodevices. It brings new insights into the field of time-dependent (non-adiabatic) quantum thermoelectricity but also suffers from severe limitations of the employed model. Indeed, the presence of a finite displacement current in our non-interacting RLM plays a crucial role in the behavior of the thermodynamic efficiencies in Sec.\ref{sec_eff} and this picture will be drastically modified after inclusion of electron-electron interactions. Moreover, it is obviously delicate to define the efficiency of a driven heat engine without including the electrostatic environment of the junction (\textit{i.e.} the load).  Both ingredients could be included in future numerical simulations, either with a self-consistent mean-field approach or with more advanced techniques.

\acknowledgments
We acknowledge the financial support of the Cross-Disciplinary Program on Numerical Simulation of CEA, the French Alternative Energies and Atomic Energy Commission. We thank Christophe Goupil for bringing Ref.\cite{Snyder2002} to our attention.

\section*{Data availability}
The data that support the findings of this study are available from the corresponding author upon reasonable request.

\appendix

\section{Derivation of generic formulas in the wide-band limit}
\label{appendixA}

We outline here the derivation of Eqs.\eqref{eq_IN(t)_1}-\eqref{eq_P_general_WBL}  and \eqref{eq_Tdyn}-\eqref{eq_DeltaQ_WBL} obtained in the WBL in Section \ref{sec_analytics_WBL}. The WBL hypothesis has two main practical consequences. First, $k(E)$ and $v(E)$ can be approximated by their values $k$ and $v$ at $E=0$. This allows us to rewrite the scattering states in Eq.\eqref{eq_PsiLE} as
\begin{subequations}
\label{eq_PsiLE_WBL}
\begin{align}
    \Psi^{LE}_{n<0}(t) &= \tfrac{1}{\sqrt{|v|}}[e^{-iEt+ikn}+e^{-ikn} r(t,E)] \\
    \Psi^{LE}_{n>0}(t) &= \tfrac{1}{\sqrt{|v|}}e^{ikn}d(t,E)
\end{align}
\end{subequations}
with $e^{ikn}/\sqrt{|v|}\approx i^n/\sqrt{2\lambda\gamma}$, and similarly for $\Psi^{RE}_{n\neq0}(t)$ with $r'(t,E)$ and $d'(t,E)$.\\
\indent Second, the scattering amplitudes can be evaluated in the WBL by doing a gauge transformation to move the time dependent pulse $V(t)$ from the dot to the leads, and then by combining the scattering amplitudes of three elementary scatterers in series (the pulse in the left lead, the time-independent dot, and the pulse in the right lead). To proceed, we introduce the generic notation $S_{\alpha\beta}$ for the scattering amplitudes, with $S_{LL}=r$,  $S_{RR}=r'$,  $S_{RL}=d$, and  $S_{LR}=d'$. In the stationary case (\textit{i.e.} when $V(t)=0)$, $S_{\alpha\beta}(t,E)=e^{-iEt}S_{\alpha\beta}^0(E)$ with likewise $S_{LL}^0=r_0$,  $S_{RR}^0=r_0'$,  $S_{RL}^0=d_0$, and  $S_{LR}^0=d_0'$. The stationary scattering amplitudes read
\begin{subequations}
\label{eqs_r0d0}
\begin{align}
    d_0(E)&=\frac{\sqrt{\Gamma_L\Gamma_R}}{i(V_0-E)+\frac{\Gamma_L+\Gamma_R}{2}}  \label{eq_d0(E)}\\
    d_0'(E)&= d_0(E) \\
    r_0(E)&= \frac{-i(V_0-E)+\frac{\Gamma_L-\Gamma_R}{2}}{i(V_0-E)+\frac{\Gamma_L+\Gamma_R}{2}} \\
    r_0'(E)&= \frac{-i(V_0-E)+\frac{\Gamma_R-\Gamma_L}{2}}{i(V_0-E)+\frac{\Gamma_L+\Gamma_R}{2}}\,. 
\end{align}
\end{subequations}
They satisfy $|r_0|^2=|r'_0|^2=1-|d_0|^2$. To calculate the scattering amplitudes in the time-dependent case, we start from the RLM model sketched in Fig.\ref{fig_RLM_sys} and make a gauge transformation to move the time dependency into the leads. Then we assume that due to the WBL approximation, we can restrict the time dependency to the outermost parts of the lead (\textit{e.g.} the left lead does not stop at site $-1$ but at site $-i$ with large $i$). The scattering problem can now be solved by combining the scattering amplitudes of a voltage pulse in an infinite lead and of the stationary dot. Importantly, in the WBL approximation, electrons are perfectly transmitted across the abrupt voltage drops in the leads (no reflections) but their energies are redistributed and we have  $d_p(E',E)=K^*(E-E')$ and $d'_p(E',E)=K(E'-E)$, where $d_p(E',E)$ and $d'_p(E',E)$ are respectively the left-to-right and right-to-left transmission amplitudes associated to the pulse in the right lead, or by symmetry the right-to-left and left-to-right transmission amplitudes associated to the pulse in the left lead. Here $K(U)=\int\!\mathrm{d}t\,e^{iUt}e^{i\phi(t)}$ with $\phi(t)=\int_{0}^t\!\mathrm{d}u\,V(u)$. Therefore\cite{gaury2014a}
\begin{equation}
    \label{eq_app_S(E',E)}
    S_{\alpha\beta}(E',E)=\int\!\frac{d\varepsilon}{2\pi}\, d_p(E',\varepsilon)S_{\alpha\beta}^0(\varepsilon)d_p'(\varepsilon,E)\,.
\end{equation}
Since $d_0(E)=d_0'(E)$, $d(t,E)=d'(t,E)$ but in general $r(t,E)\neq r'(t,E)$. We find 
\begin{subequations}
\label{eq_dd'rr'_WBL}
\begin{align}
  d(t,E) & =e^{-i\phi(t)}\int\!\frac{\mathrm{d}E'}{2\pi}d_0(E')K(E'-E)e^{-iE't} \label{eq_d(t,E)}\\
  d'(t,E) & =d(t,E) \label{eq_d'_WBL}\\
  r(t,E)&=\sqrt{\tfrac{\Gamma_L}{\Gamma_R}}\,d(t,E)-e^{-iEt} \label{eq_r_WBL}\\
  r'(t,E)&=\sqrt{\tfrac{\Gamma_R}{\Gamma_L}}\,d'(t,E)-e^{-iEt}\,.  \label{eq_r'_WBL}
\end{align}
\end{subequations}
Eq.\eqref{eq_r_WBL} linking $d(t,E)$ and $r(t,E)$ can be easily derived by writing down the time-dependent Schr\"odinger equation for the scattering states \eqref{eq_PsiLE} upon neglecting the energy dependency of the velocity $v(E)$ (WBL approximation). And similarly for Eq.\eqref{eq_r'_WBL}. Eq.\eqref{eq_d(t,E)} is also trivial to show by writing the Fourier transform of Eq.\eqref{eq_app_S(E',E)}. By inserting Eqs.\eqref{eq_PsiLE_WBL} and \eqref{eq_dd'rr'_WBL} into Eqs.\eqref{eq_IN(t)_general}, \eqref{eq_IH(t)_general} and \eqref{eq_P_df}, we deduce Eqs.\eqref{eq_IN(t)_1}-\eqref{eq_P_general_WBL}.\\
\indent Finally, it can be shown using Eqs.\eqref{eq_d(t,E)} and \eqref{eq_d0(E)} that $d(t,E)$ obeys 
\mathleft
\begin{subequations}
\label{eqs_properties_d}
\begin{align}
    \partial_t|d|^2 &= -2\Gamma\,|d|^2+2\sqrt{\Gamma_L \Gamma_R}\,\mathrm{Re}\left[e^{iEt}d\right] \label{eq_property1_d} \\
    \mathrm{Im}\left[d^*\partial_t d\right] &= - \epsilon_0(t)|d|^2-\sqrt{\Gamma_L \Gamma_R}\,\mathrm{Im}\left[e^{iEt}d\right] \\
    \mathrm{Im}\left[e^{iEt}\partial_t d\right] &= - \epsilon_0(t)\mathrm{Re}\left[e^{iEt}d\right]-\Gamma\,\mathrm{Im}\left[e^{iEt}d\right]
\end{align}
\end{subequations}
\mathcenter
where $\Gamma=(\Gamma_L+\Gamma_R)/2$ and $d$ is a shorthand notation for $d(t,E)$. Moreover,
\begin{flalign}
   \int\!\frac{\mathrm{d}E}{2\pi}|d(t,E)|^2=\!\int\!\frac{\mathrm{d}E}{2\pi}[1\!-\!|r(t,E)|^2]=\frac{\Gamma_L\Gamma_R}{\Gamma_L+\Gamma_R} 
   \label{eq_csv_dr}
\end{flalign}
and the same equation holds by replacing $r$ by $r'$. Using Eqs.\eqref{eq_d'_WBL}-\eqref{eq_r_WBL}, \eqref{eq_property1_d}, and the fact that $V(t)$ is supposed to be finite in time in Section \ref{sec_timeintegrated_WBL} (so that $\int\!\mathrm{d}t\, \partial_t |d|^2=0$), we deduce Eqs.\eqref{eq_Tdyn} and \eqref{eq_Tdyn2}. The fact that $\int\!\mathrm{d}E\,T_{dyn}(E)=0$ follows from Eq.\eqref{eq_csv_dr}. Then Eqs.\eqref{eq_IN(t)_1}-\eqref{eq_IN(t)_3} give Eq.\eqref{eq_DeltaN}. Finally, using on one hand Eqs.\eqref{eq_IH(t)_WBL}, \eqref{eq_Tdyn}-\eqref{eq_DeltaN} to express $\Delta Q_\alpha$, and on the other hand the set of equations \eqref{eqs_properties_d} and Eq.\eqref{eq_P_general_WBL} to express $W_{ext}$, as well as again the fact that $V(t)$ is finite in time, we derive Eq.\eqref{eq_DeltaQ_WBL}.

\section*{References}
\bibliography{biblio}

\end{document}